\documentclass[12pt,a4paper,final]{iopart}

\usepackage{iopams}
\usepackage{graphicx}
\usepackage{subfigure}
\usepackage[breaklinks=true,colorlinks=true,linkcolor=blue,urlcolor=blue,citecolor=blue]{hyperref}
\usepackage{cite}

	\newcommand{\ket}[1]{\left| #1 \right\rangle}
	\newcommand{\bra}[1]{\left\langle #1 \right|}
	
	\newcommand{\braket}[2]{\left\langle #1 \right. \left| #2 \right\rangle}

\begin{document}

\title[Dynamics and Protecting of Entanglement...]{Dynamics and Protecting of Entanglement in Two-Level Systems Interacting with a Dissipative Cavity: The Gardiner-Collett Approach}

\author{A Nourmandipour$^{1}$}
\address{$^1$Atomic and Molecular Group, Faculty of Physics, Yazd University, Yazd  89195-741, Iran}
\ead{anoormandip@stu.yazd.ac.ir}

\author{M K Tavassoly$^{1,2}$}
\address{$^1$Atomic and Molecular Group, Faculty of Physics, Yazd University, Yazd  89195-741, Iran
\\ $^2$Photonic Research Group, Engineering Research Center, Yazd University, Yazd  89195-741, Iran}
\ead{mktavassoly@yazd.ac.ir}

\begin{abstract}
In this paper, we study the exact entanglement dynamics of two two-level atoms in a dissipative cavity. We use the Gardiner-Collett Hamiltonian to model the dissipative cavity, in which, we assume that the two atoms resonantly interact with the cavity field and the cavity field itself interacts with the surrounding medium. Then, with the help of the Fano's technique, we show that, this system can be regarded as two atoms interacting with a heat bath. In such a case, we find that, there exists a decoherence-free state that does not evolve in time. At this time, there exists a so-called super-radiant state which decays in time due to dissipation. At last, we use the quantum Zeno effect to preserve the entanglement which already has been stored in the system.
\end{abstract}

\vspace{2pc}

\submitto{J. Phys. B}

\maketitle

\section{Introduction}\label{Introduction}

Entanglement is a pure quantum mechanical phenomenon which shows the quantum correlations between multi sub-systems. This strange phenomenon has recently attracted a great deal of attention \cite{Bennett1992}. This notion has many applications such as quantum cryptography \cite{Ekert1991}, quantum teleportation \cite{Braunstein1995}, superdense coding \cite{Mattle1996}, entanglement swapping \cite{Hu2011}, sensitive measurements \cite{Richter2007} and quantum telecloning \cite{Muarao1999}. There are many implementations to produce entangled states, such as trapped ions \cite{Turchette1998}, atomic ensembles \cite{Julsgaard2001}, photon pairs \cite{Aspect1981} and superconducting qubits \cite{Izmalkov12004}. It is well-known that the interaction of atoms with various types of cavity field (with additional interaction terms such as Kerr medium, etc.) is an efficient source of entanglement \cite{Faghihi2013}. On the other hand, because of the unavoidably interaction between any real system with its surrounding environment, dissipation is ever present. This natural process usually leads to loss of entanglement stored in those systems, i.e. disentanglement occurs. So, many attentions have been paid to the theory of open quantum systems \cite{Breuer2002}. In this regard, the Lindblad master equation can be used to express the quantum master equation for a dissipative system which is based on the temporal evolution of the density operator of the system \cite{Breuer2002,Memarzadeh2013}. In a different way, one can deal with the time evolution of the wave function of the system instead of density operator by solving the time-dependent Schr\"{o}dinger equation. Recently, the case of two two-level atoms dissipating into a common heat bath has been considered in \cite{Maniscalco2008}, where the authors have considered the surrounding environment as a zero-temperature bosonic reservoir and solved the related time dependent Schr\"{o}dinger equation. In another point of view, one of the interesting topics in dissipative systems is to find a way to fight against the deterioration of the entanglement. Many schemes have been proposed in order to preserve entanglement in such systems in literature. For instance, it has been shown that, addition of a laser field leads to the high stationary entanglement \cite{Soltani2013}. Another approach to overcome this problem, relies on active feedback \cite{Sayrin2011}. Beside these, the quantum Zeno effect (QZE) is a promising way to avoid the decaying behaviour of the entanglement in dissipative systems. This effect which refers to the inhibition of the temporal evolution of a quantum system by repeated projective measurements during a defined period of time, has been discussed theoretically \cite{Misra1977} as well as experimentally \cite{Sayrin2011}.

In this paper, we intend to extend the idea of a dissipative cavity into a model, in which the atoms interact with a cavity field and the cavity mode itself interacts with the surrounding environment. We model the surrounding environment as a set of continuum harmonic oscillators. In this regard, this idea can be used for modelling the dissipative cavity that photons in the cavity should leak out to a continuum of states. The Hamiltonian describing this model of dissipation is called the Gardiner-Collett Hamiltonian $(H_{\mathrm{GC}})$. This model leads to a Lorentzian spectral density which is directly obtained from our modelling of the dissipative cavity. Whereas, in the previous studies, it is assumed that the spectral density should be a Lorentzian one \cite{Maniscalco2008}.  This kind of spectral density implies the nonperfect reflectivity of the cavity mirrors. We obtain the exact entanglement dynamics as a function of the environment correlation time for both weak and strong couplings corresponding to the bad and good cavity limits, respectively. It should be noticed that, our results are outside of the Markovian regime. In fact, for special values of the environment correlation time, we are able to obtain the two-atom Jaynes-Cummings model and Markovian regime. For typical values of this parameter, our model interpolates between these two limits. In addition, considering the QZE, we use the action of a series of nonselective measurements on the collective atomic system, showing that, the presence of measurements can quenches the decay of the entanglement. Our results are verifiable and confirmable with a slight modification in both cavity QED experiments with atoms (or ions) trapped in an electromagnetic cavity \cite{Guthohilein2001} and superconducting Josephson circuits \cite{Wallraff2004}.

The rest of the paper is organized as follow: In section 2, we introduce the Hamiltonian of the system and simplify it by using the Fano's technique. In section 3 we try to find the wave vector of the system. Section 4 deals with the investigation of the entanglement dynamics of the system by various measures. In section 5, we introduce the quantum Zeno effect and show how to preserve the entanglement. Finally, the results are summarized in section 6.

\section{Model}
The system under our consideration consists of a dissipative cavity contains two two-level atoms with excited (ground) state $\ket{e}$ ($\ket{g}$). Because of the presence of dissipation, the Gardiner-Collett approach to describe the dissipation seems to be useful. The Gardiner-Collett Hamiltonian for a dissipative cavity is written as $(\hbar=c=1)$ \cite{Collet1984,Dalton2001,Dalton2003}

\begin{equation}
 \hat{H}_{\mathrm{GC}}= \omega_c \hat{a}^{\dagger} \hat{a} +  \int_0^{\infty}\! \eta \hat{B}^{\dagger}(\eta)\hat{B}(\eta) \, \mathrm{d}\eta +  \int_0^{\infty}\! \left\lbrace  G(\eta)\hat{a}^{\dagger}\hat{B}(\eta)\ + \ \mathrm{H.c.} \right\rbrace \, \mathrm{d}\eta,
  \label{eq:gardinerhamiltonian}
\end{equation}
where $\hat{a}$ ($\hat{a}^{\dagger}$) and $\omega_c$ are the annihilation (creation) and frequency of the cavity field, respectively. $G(\eta)$ is the coupling coefficient which in general, is a function of frequency that connects the external world to the cavity, and $\hat{B}^{\dagger}(\eta)$ and $\hat{B}(\eta)$ are the creation and annihilation operators of the surrounding environment at mode $\eta$ which obey the following commutation relation
 \begin{equation}
 \left[ \hat{B}(\eta),\hat{B}^{\dagger}(\eta^{'}) \right]  =\delta(\eta-\eta^{'}).
 \label{eq:commutationrelationsurrounding}
 \end{equation}
 The second term of (\ref{eq:gardinerhamiltonian}) can be interpreted as the Hamiltonian of a driving single-excitation source \cite{Razavi2006}.
 Altogether, the suitable Hamiltonian which describes our system is written as
 \begin{equation}
\hat{H}=\hat{H}_\mathrm{A}+ \hat{H}_{\mathrm{FE}} + \hat{H}_{\mathrm{AF}},
\label{eq:H}
  \end{equation}
where
\numparts
\begin{eqnarray}
\hat{H}_\mathrm{A}&=\sum_{i=1}^{2}\omega_{i}\hat{\sigma}_{+}^{(i)}\hat{\sigma}_{-}^{(i)}, \label{eq:HA} \\
\hat{H}_{\mathrm{FE}}&=\omega_{c}\hat{a}^{\dagger}\hat{a} + \int_0^{\infty}\! \eta \hat{B}^{\dagger}(\eta)\hat{B}(\eta) \, \mathrm{d}\eta
+ \int_0^{\infty}\! \left(   G(\eta)\hat{a}^{\dagger}\hat{B}(\eta)\ + \ \mathrm{H.c.} \right)  \, \mathrm{d}\eta, \label{eq:HFE} \\
\hat{H}_{\mathrm{AF}}&= g_1\hat{\sigma}_+^{(1)}\hat{a}+g_2\hat{\sigma}_+^{(2)}\hat{a} + \mathrm{H.c.} \label{eq:HAF}
\end{eqnarray}
\endnumparts
In the above relation, $\hat{H}_\mathrm{A}$ is the Hamiltonian of the atoms, $\hat{H}_{\mathrm{FE}}$ represents the Hamiltonian of the cavity fields, the surrounding environments as well as their interaction, and $\hat{H}_{\mathrm{AF}}$ denotes the interaction between atoms and the cavity field. In the above set of relations $\sigma_{+}^{(i)}$ ($\sigma_{-}^{(i)}$) is the raising (lowering) operator of the $i$th atom, $\omega_{i}$ is the resonance frequency of the $i$th atom and $g_{i}$ is the coupling constant between $i$th atom and the cavity field.
In the continuation, we assume that the surrounding environment has such a narrow bandwidth such that only a particular mode of the cavity may be excited \cite{Dutra2005}. This assumption allows us to extend integrals over $\eta$ back to $-\infty$ and to take $G(\eta)$ as a constant (equal to $\sqrt{\kappa/\pi}$).
The Hamiltonian (\ref{eq:HFE}) can now be diagonalized using Fano's technique \cite{Fano1961,Barnett1997}. This approach consists in finding the set of annihilation and creation operators that diagonalizes the $H_{\mathrm{FE}}$ Hamiltonian. To achieve this purpose, let us define the dressed operator
\begin{equation}
\hat{A}(\omega)=\alpha(\omega)\hat{a}+\int\! \beta(\omega,\eta)\hat{B}(\eta) \, \mathrm{d}\eta,
\label{wq:dressedoperator}
\end{equation}
where $\alpha(\omega)$ and $\beta(\omega,\eta)$ are obtained such that $\hat{A}(\omega)$ is an annihilation operator which obeys the following commutation relations with its conjugate
\begin{equation}
\left[ \hat{A}(\omega),\hat{A}^{\dagger}(\omega^{'})\right]=\delta(\omega-\omega^{'})
\label{eq:commutationrelation}
\end{equation}
and the $\hat{H}_{\mathrm{FE}}$ Hamiltonian is diagonal, i.e.
\begin{equation}
 \hat{H}_{\mathrm{FE}}=\int\! \omega \hat{A}^{\dagger}(\omega)\hat{A}(\omega) \, \mathrm{d}\omega \label{eq:digonalizationcondition2}
\end{equation}
so that
\begin{equation}
\left[ \hat{A}(\omega),\hat{H}_{\mathrm{FE}}\right] =\omega\hat{A}(\omega). \label{eq:digonalizationcondition1}
\end{equation}
After some manipulations, we have (see appendix A)
\numparts
\begin{eqnarray}
\alpha(\omega)&=\frac{\sqrt{\kappa/\pi}}{\omega-\omega_c+i\kappa}, \label{eq:alpha} \\
\beta(\omega,\eta)&= \sqrt{\kappa/\pi}\alpha(\omega)\left[  P\frac{1}{\omega-\eta}+\frac{\omega-\omega_c}{\kappa/\pi}\delta(\omega-\eta)\right], \label{eq:beta}
\end{eqnarray}
\endnumparts
where $P$ refers to the principal value.
We can thus express $\hat{a}$ in (\ref{wq:dressedoperator}) as a linear combination of $\hat{A}(\omega)$; that is (see appendix A)
\begin{equation}
\hat{a}=\int\! \alpha^{*}(\omega)\hat{A}(\omega) \, \mathrm{d}\omega.
\label{eq:operatorA}
\end{equation}
Consequently, the Hamiltonian of our system can be finally rewritten in terms of the dressed operators as follows
\begin{eqnarray}
\hat{H}&=\sum_{i=1}^{2}\omega_{i}\hat{\sigma}_{+}^{(i)}\hat{\sigma}_{-}^{(i)}+\int\! \omega \hat{A}^{\dagger}(\omega)\hat{A}(\omega) \, \mathrm{d}\omega \nonumber \\
&+ \int\!\left(\left(g_1\hat{\sigma}_+^{(1)}+g_2\hat{\sigma}_+^{(2)} \right)\alpha^{*}(\omega)\hat{A}(\omega)+\mathrm{H.c.}\right) \, \mathrm{d}\omega.
\label{eq:finalhamiltonian}
\end{eqnarray}
The obtained Hamiltonian clearly implies that, the two atoms are dissipating in a common heat bath.

\section{Time evolution of the entangled states of the system}

Now, we try to solve the time-dependent Schr\"{o}dinger equation and obtain the state vector of the system at any time $t$. Before that, we introduce the collective coupling constant as $g_{_{T}}=(g_1^2+g_2^2)^{1/2}$ and dimensionless relative strengths $r_j=\frac{g_j}{g_T}$ ($r_1^2+r_2^2=1$), in which we take only $r_1$ as independent.
We assume that there is no excitation in the cavities before the occurrence of interaction and the two atoms are in a general superposition in the following form
\begin{equation}
\ket{\psi_0}=\left( c_{01}\ket{e,g}+c_{02}\ket{g,e}\right)\ket{\boldsymbol{0}}_{R},
\label{eq:initialstate}
\end{equation}
in which $\ket{\boldsymbol{0}}_{R}=\hat{A}(\omega)\ket{1_{\omega^{'}}}\delta(\omega-\omega^{'})$ is the multi-mode vacuum state, where $\ket{1_{\omega}}$ is the multi-mode state representing one photon at frequency $\omega$ and vacuum state in all other modes.
Accordingly, the quantum state of the entire system at any time can be written as

\begin{equation}
\ket{\psi(t)}=c_1(t)\ket{e,g}\ket{\boldsymbol{0}}_{R}+c_2(t)\ket{g,e}\ket{\boldsymbol{0}}_{R}
 +\int\! c_{\omega}(t)\ket{1_{\omega}}\ket{g,g} \, \mathrm{d}\omega.
\label{eq:state}
\end{equation}

Using the time-dependent Schr\"{o}dinger equation $\left( i\dot{\ket{\psi}}=\hat{H}\ket{\psi}\right) $, one arrives at the following set of coupled integro-differential equations
\numparts
\begin{eqnarray}
\dot{u}_j(t)&=-ig_j\int\! \alpha^*(\omega) e^{i\delta_{\omega}^{(j)}t}u_{\omega}(t) \, \mathrm{d}\omega, \label{eq:diff1} \\
\dot{u}_{\omega}(t)&=-i\alpha(\omega)\left( g_1u_1(t)e^{-i\delta_{\omega}^{(1)}t}+g_2u_2(t)e^{-i\delta_{\omega}^{(2)}t} \right). \label{eq:diff2}
\end{eqnarray}
\endnumparts
In the above relations we have used
\numparts
\begin{eqnarray}
u_j(t)&=c_j(t)e^{i\omega_jt}, \label{eq:uj} \\
u_{\omega}(t)&=c_{\omega}(t)e^{i\omega t}, \label{eq:uw} \\
\delta_{\omega}^{(j)}&=\omega_j-\omega. \label{eq:delta}
\end{eqnarray}
\endnumparts
In the following, we assume that the two atoms have the same Bohr frequency, i.e. $\omega_1=\omega_2=\omega_0$, consequently, $\delta_{\omega}^1=\delta_{\omega}^2=\delta_{\omega}\equiv\omega_0-\omega$, and also we assume that the two atoms interact with cavity field in the exact resonance condition, i.e. $\omega_0-\omega_c=0$. By integrating Eq. (\ref{eq:diff2}) and inserting its solution into Eq. (\ref{eq:diff1}), one obtains two intero-differential equations for amplitudes $u_{1,2}(t)$ as follow
\numparts
\begin{eqnarray}
\dot{u}_{1}(t)&=-\int_{0}^{t}\! f(t-t^{'})\left( g_1^2u_1(t^{'})+g_1g_2u_2(t^{'})\right) \, \mathrm{d}t^{'}, \label{eq:diffu1} \\
\dot{u}_{2}(t)&=-\int_{0}^{t}\! f(t-t^{'})\left( g_2^2u_2(t^{'})+g_1g_2u_1(t^{'})\right) \, \mathrm{d}t^{'}, \label{eq:diffu2}
\end{eqnarray}
\endnumparts
in which the correlation function $f(t-t^{'})$ reads as
\begin{equation}
f(t-t^{'})=\int\! \left| \alpha(\omega)\right| ^2e^{i\delta_{\omega}(t-t^{'})} \, \mathrm{d}\omega. \label{eq:f}
\end{equation}
In this regard and according to Eq. (\ref{eq:alpha}), one can easily observe that $\left| \alpha(\omega)\right| ^2=\kappa/ \pi\left[ (\omega-\omega_c)^2+\kappa^2\right] $ is a Lorentzian spectral density which implies the nonperfect reflectivity of the cavity mirrors. Note that, this result is directly obtained from our modelling of dissipative cavity. In this case, the correlation function decays exponentially $f(t-t^{'})=e^{-\kappa (t-t^{'})}$, where $\kappa$ being the decay rate factor of the cavity. Consequently, the quantity $1/\kappa$ is the cavity correlation time. For an ideal cavity (i.e. $\kappa\rightarrow 0$), $\left| \alpha(\omega)\right| ^2=\delta(\omega-\omega_c)$ corresponds to a constant correlation function. In this situation, the system reduces to a two-atom Jaynes-Cummings model \cite{Tavis1968} with vacuum Rabi frequency $\Omega_R=g_{_{T}}$. On the other hand, in the Markovian regime, i.e., for small correlation times (with $\kappa$ much larger than any other frequency scale), we obtain the decay rate as $\gamma=2g_{_{T}}^2/\kappa$.

\section{Dynamics of Entanglement}
Using Eq. (\ref{eq:state}), the explicit form of the $4\times 4$ density matrix for atoms in the $\left\lbrace \ket{e,e},\ket{e,g}, \ket{g,e}, \ket{g,g}\right\rbrace $ basis can be derived as
\begin{equation}
 \rho(t)= \left( \begin{array}{cccc}
0 & 0 & 0 & 0 \\
0 & \left| c_1(t)\right| ^2 & c_1(t)c_2^{*}(t) & 0 \\
0 &  c_1^{*}(t)c_2(t) &  \left| c_2(t)\right| ^2 &0 \\
0 & 0 & 0 & 1-\left| c_1(t)\right| ^2-\left| c_2(t)\right|^2 \end{array} \right).
\label{eq:densitymatrix}
\end{equation}
A suitable measure for degree of entanglement (DEM) for bipartite systems is concurrence. For qubits, Wootterrs defined concurrence using Puali matrix $\hat{\sigma}_{y}$ as follows \cite{Wootters1998}
\begin{equation}
C(t)=\max\left\lbrace 0, \sqrt{\lambda_1}- \sqrt{\lambda_2}- \sqrt{\lambda_3}- \sqrt{\lambda_4}\right\rbrace,
\label{eq:con}
\end{equation}
where $\left\lbrace \lambda_i\right\rbrace _{i=1}^4$ are the eigenvalues (in decreasing order) of the Hermitian matrix $\hat{R}=\hat{\rho}\hat{\rho}_s$, in which $\hat{\rho}$ is the density matrix of the system and $\hat{\rho}_s=\hat{\sigma}_y\otimes\hat{\sigma}_y\hat{\rho}^{*}\hat{\sigma}_y\otimes\hat{\sigma}_y$ where $\rho^*$ is complex conjugate of $\rho$ in computational basis. The concurrence varies between 0 (when the atoms are separable) and 1 (when they are maximally entangled). For the density matrix given by (\ref{eq:densitymatrix}), the concurrence becomes
\begin{equation}
C(t)=2\left| c_1(t)c_2^*(t)\right|.  \label{eq:concurrence}
\end{equation}

\begin{figure}[ht]
\centering
\subfigure[\label{fig:stacon0} \ $\varphi=0. $]{\includegraphics[width=0.4\textwidth]{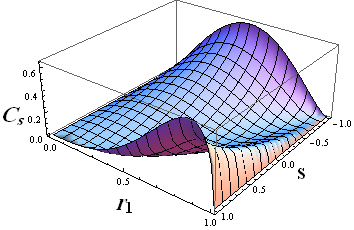}}
\hspace{0.05\textwidth}
\subfigure[ \label{fig:staconp} \ $\varphi=\pi.$]{\includegraphics[width=0.4\textwidth]{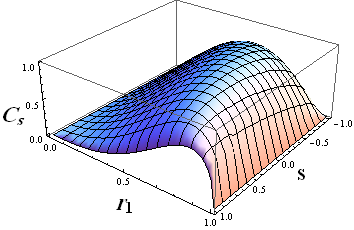}}

\caption{\label{fig:stationaryconcurrence} Stationary concurrence as a function of the relative coupling constant $r_1$ and the initial separability parameter $s$ for (a) $\varphi=0$ and (b) $\varphi=\pi$.}
\end{figure}

\begin{figure}[htb!]
\centering
\subfigure[\label{fig:bads10} \ Factorized initial state, $s=1$.]{\includegraphics[width=0.4\textwidth]{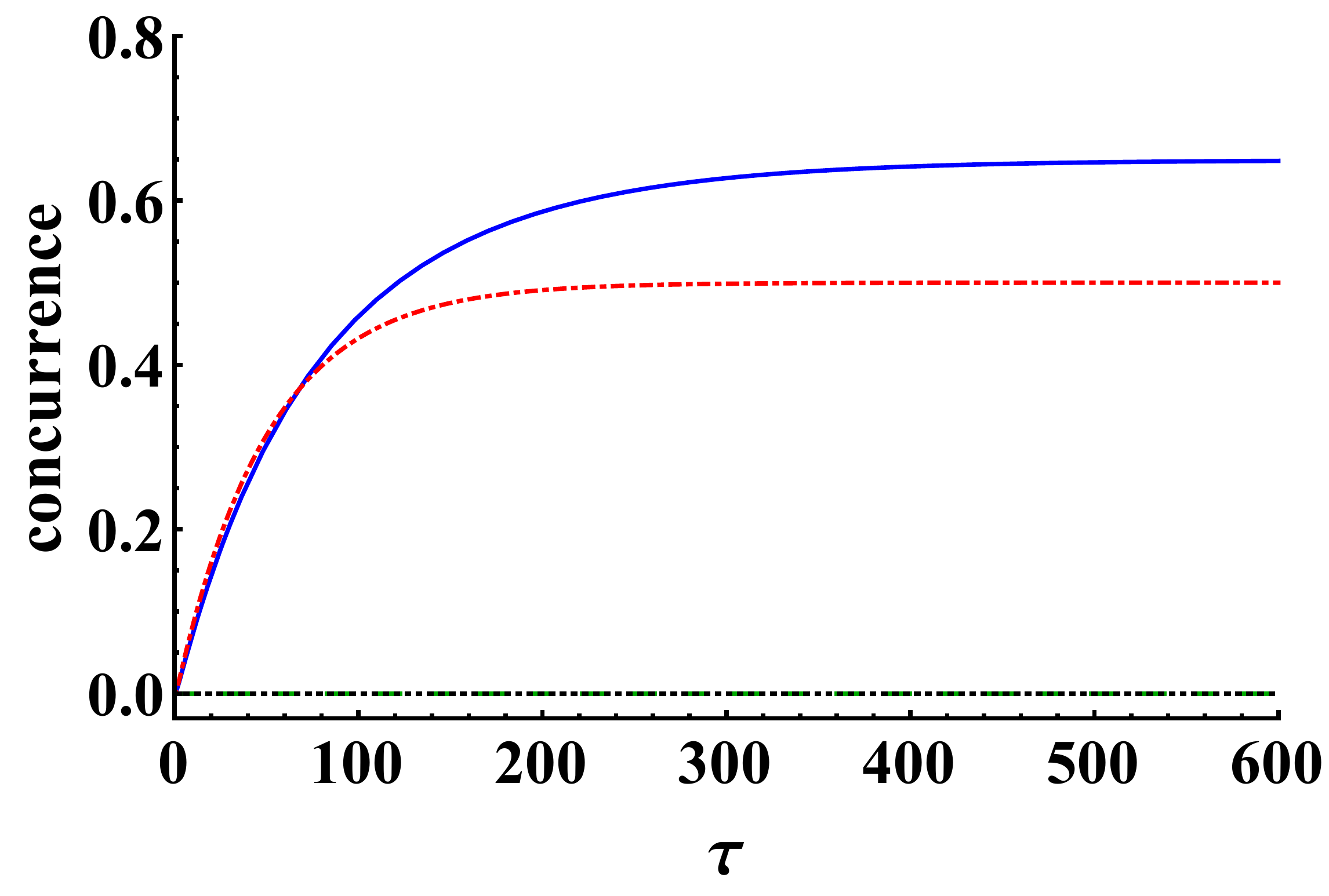}}
\hspace{0.05\textwidth}
\subfigure[\label{fig:bads00} \ Entangled initial state, $s=0$.]{\includegraphics[width=0.4\textwidth]{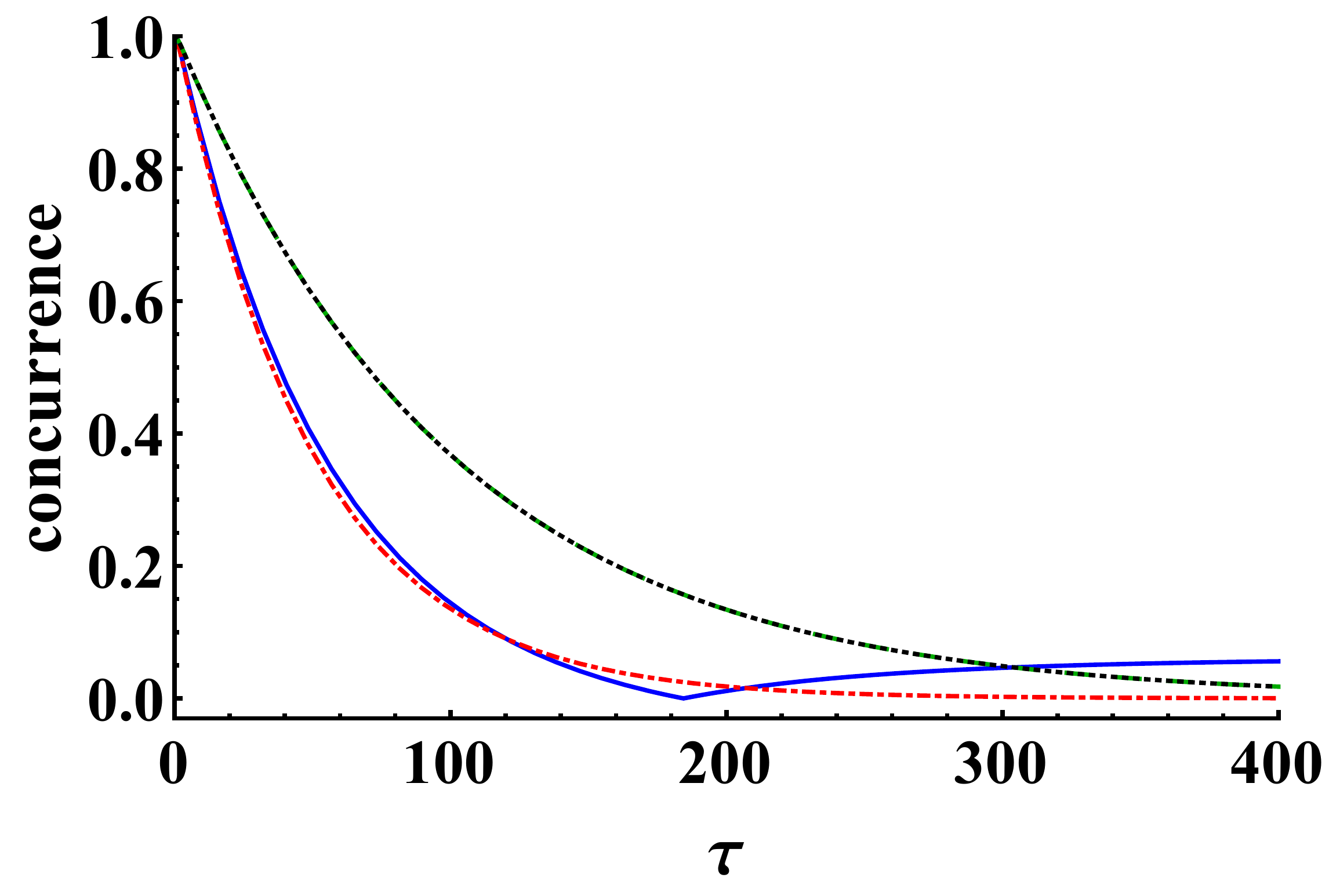}}
  \hspace{0.05\textwidth}
\subfigure[\label{fig:bads1p} \ Factorized initial state, $s=1$.]{\includegraphics[width=0.4\textwidth]{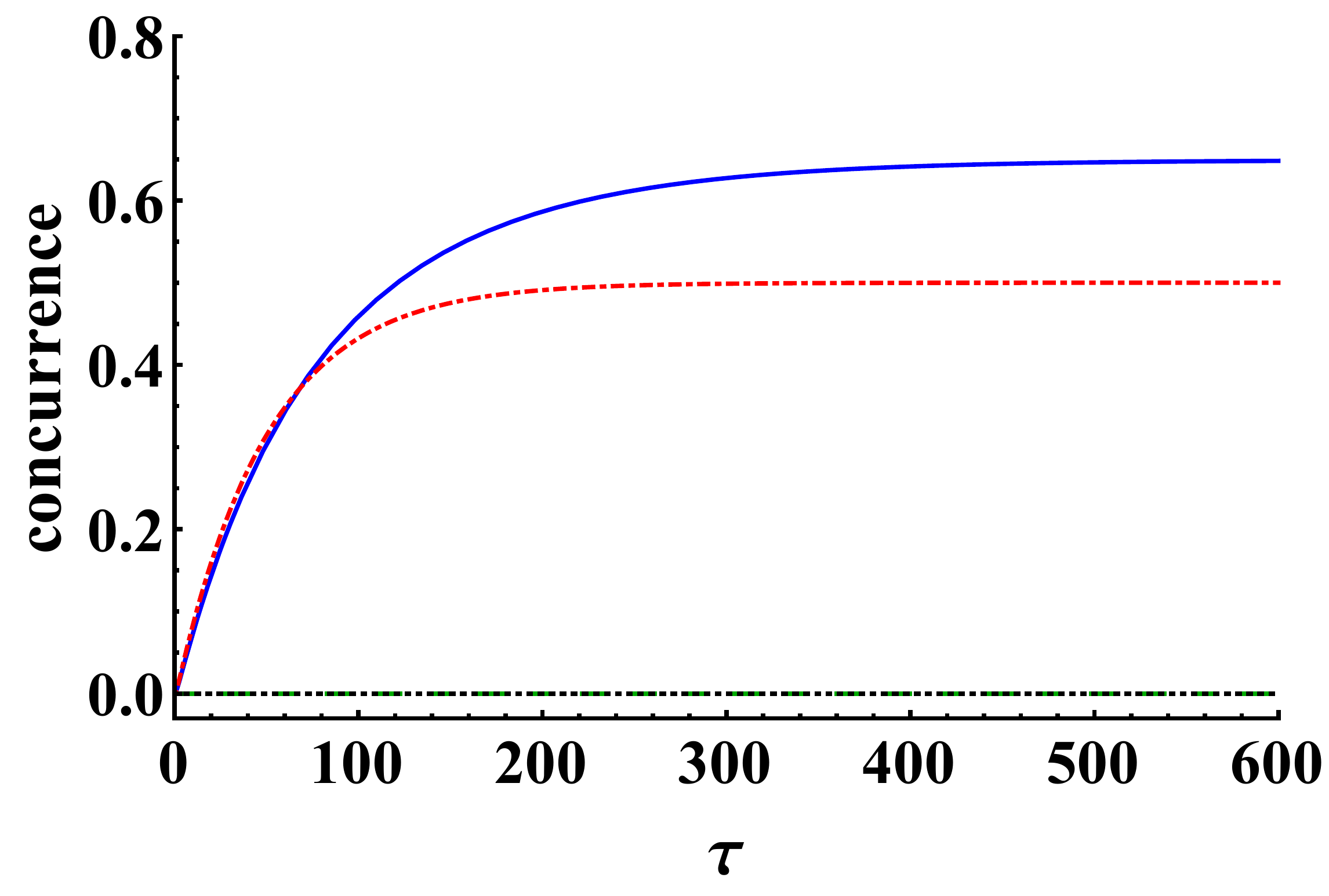}}
\hspace{0.05\textwidth}
\subfigure[\label{fig:bads0p} \ Entangled initial state, $s=0$.]{\includegraphics[width=0.4\textwidth]{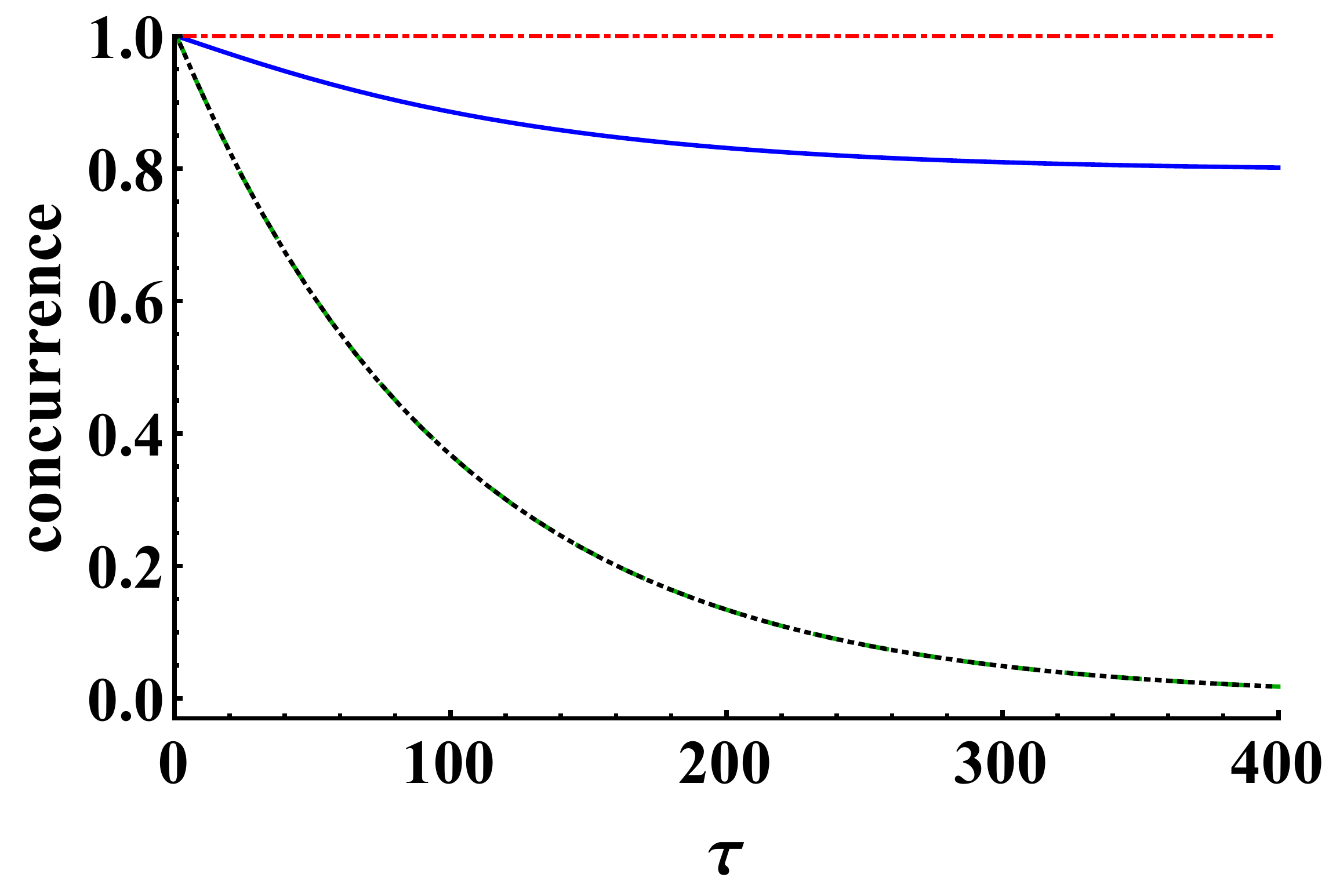}}

\caption{Concurrence as function of $\tau$ in the bad cavity limit, i.e. $R=0.1$ for $\varphi=0$ (top plots) and $\varphi=\pi$ (bottom plots) with $s=1$ (left plots) and $s=0$ (right plots) with the cases ($i$) maximal stationary value, $r_1=0.87$ (solid line), ($ii$) symmetric coupling, $r_1=1/\sqrt 2$ (dot-dashed line), and ($iii$) only one coupled atom, $r_1=0$ (dashed line) and $r_1=1$ (dotted line).} \label{fig:conbadcavity}
   \end{figure}

\begin{figure}[htb!]
\centering
\subfigure[\label{fig:goods10} \ Factorized initial state, $s=1$.]{\includegraphics[width=0.4\textwidth]{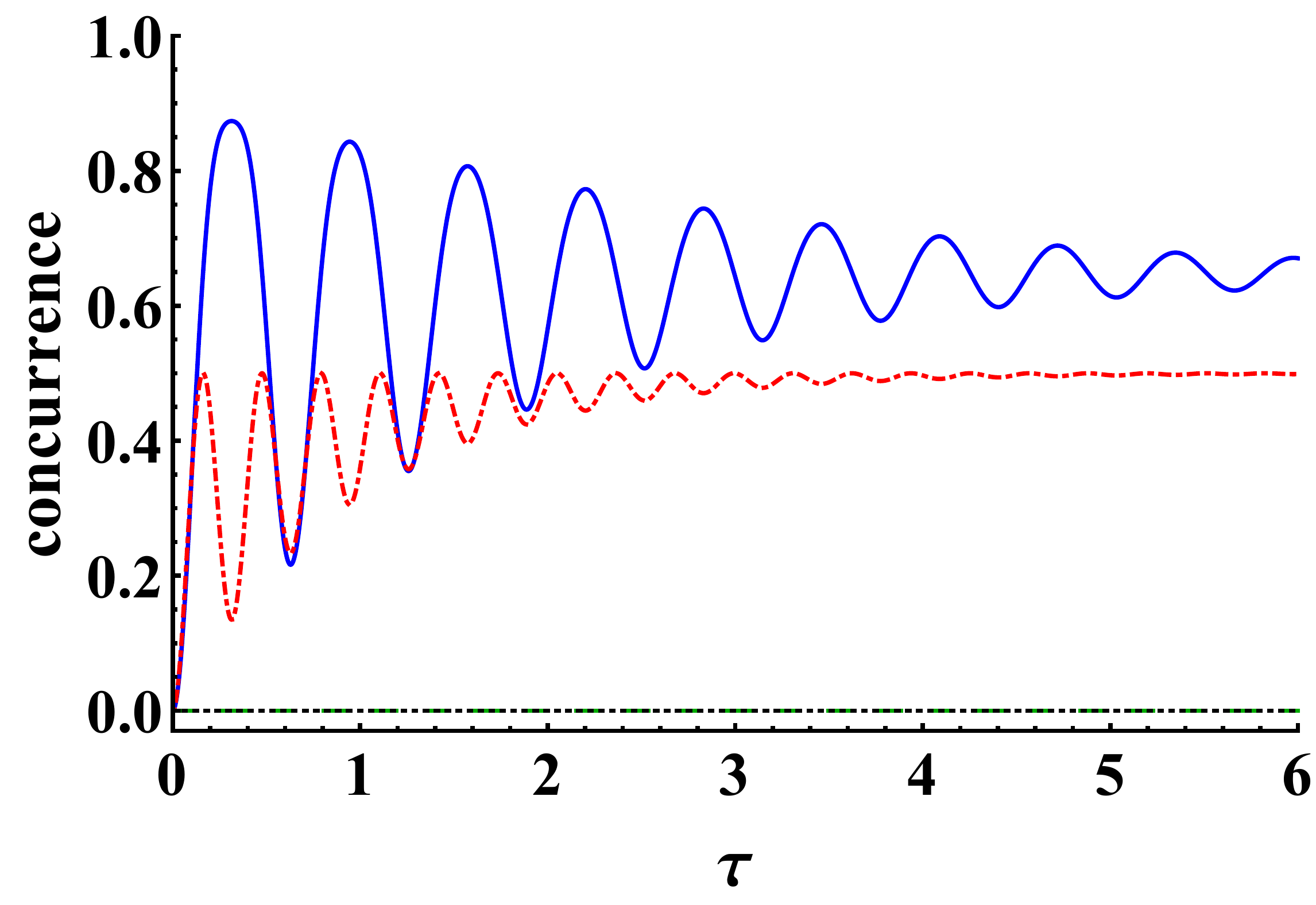}}
  \hspace{0.05\textwidth}
\subfigure[\label{fig:goods00} \ Entangled initial state, $s=0$.]{\includegraphics[width=0.4\textwidth]{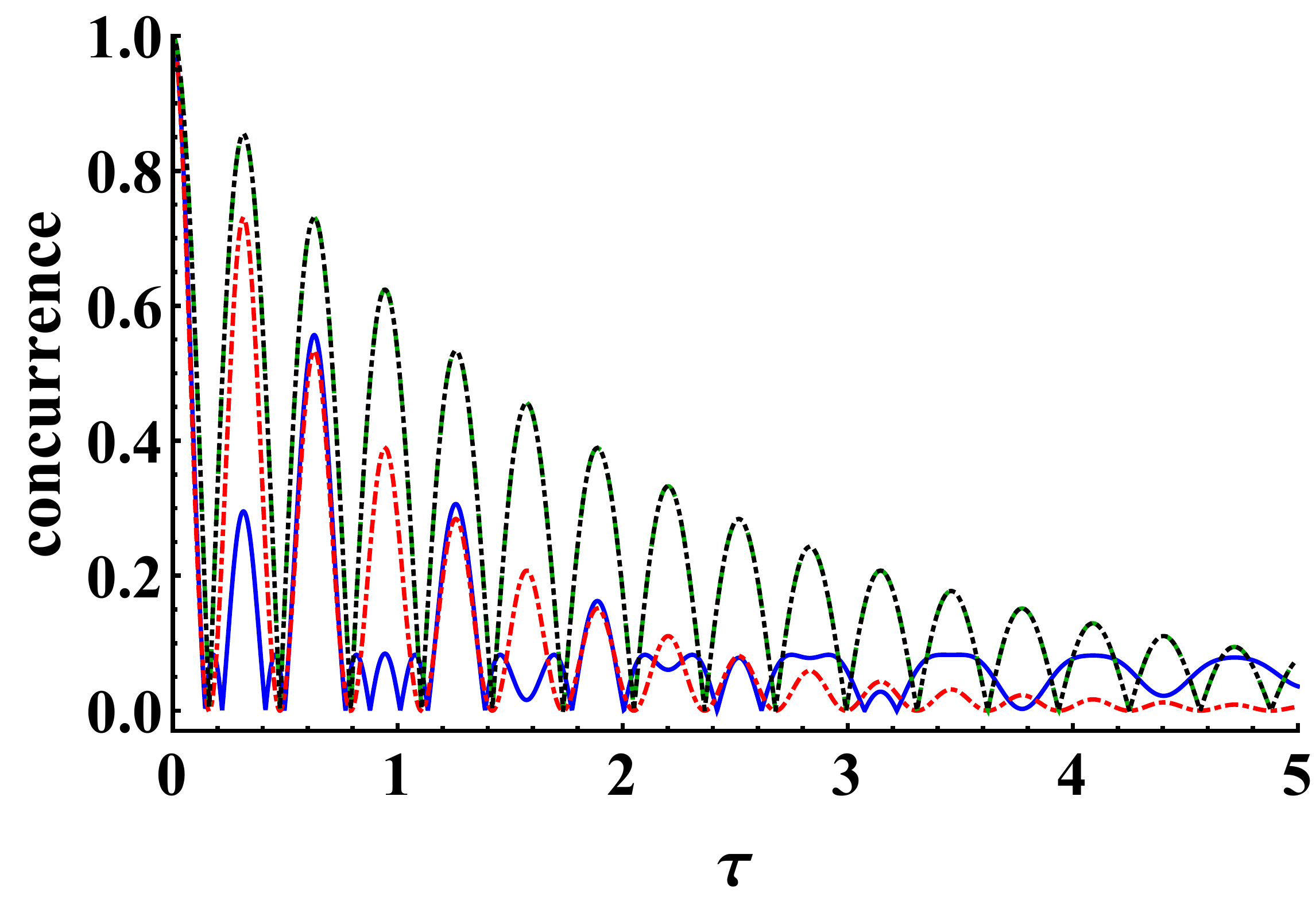}}
  \hspace{0.05\textwidth}
\subfigure[\label{fig:goods1p} \ Factorized initial state, $s=1$.]{\includegraphics[width=0.4\textwidth]{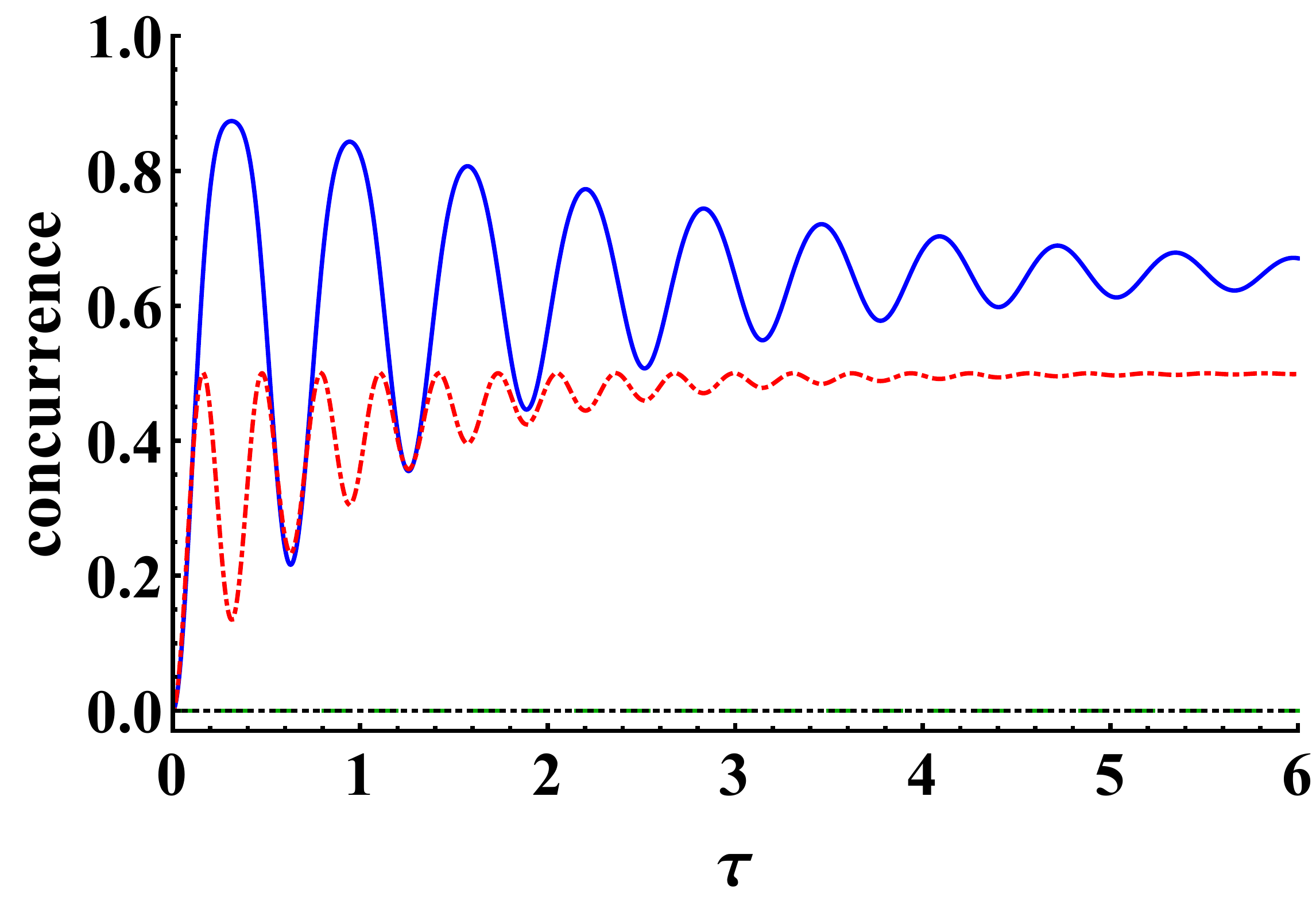}}
  \hspace{0.05\textwidth}
\subfigure[\label{fig:goods0p} \ Entangled initial state, $s=0$.]{\includegraphics[width=0.4\textwidth]{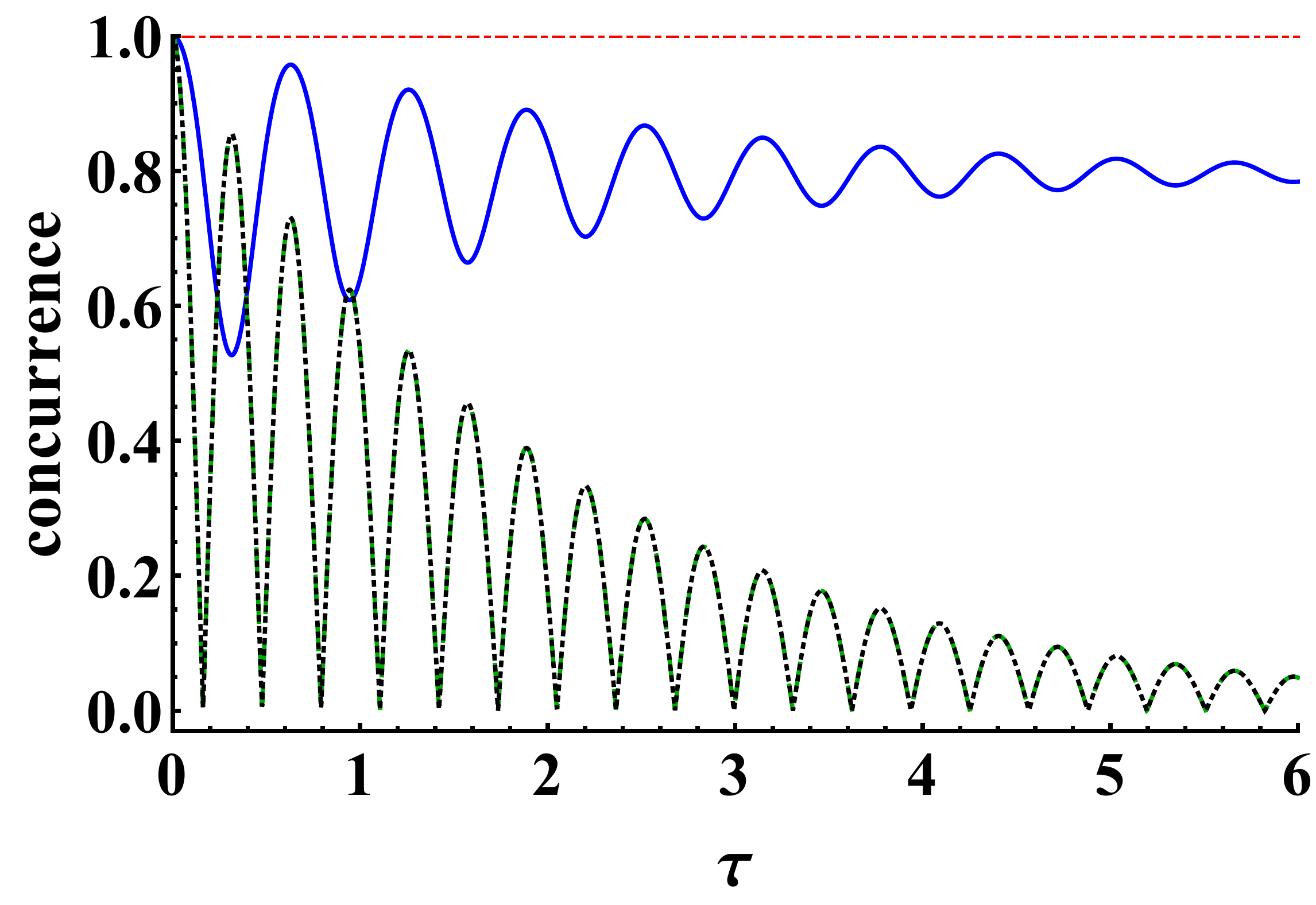}}

\caption{Concurrence as function of $\tau$ in the good cavity limit, i.e. $R=10$ for $\varphi=0$ (top plots) and $\varphi=\pi$ (bottom plots) with $s=1$ (left plots) and $s=0$ (right plots) with the cases ($i$) maximal stationary value, $r_1=0.87$ (solid line), ($ii$) symmetric coupling, $r_1=1/\sqrt 2$ (dot-dashed line), and ($iii$) only one coupled atom, $r_1=0$ (dashed line) and $r_1=1$ (dotted line).} \label{fig:congoodcavity}
   \end{figure}

It is shown that the interaction with a \emph{common} environment can generate a highly entangled long-living decoherence-free (or sub-radiant) state \cite{Maniscalco2008,Palma1996,Zanardi1997}. Consequently, before discussing the general time evolution, we intend to find such a stationary state. According to Eqs. (\ref{eq:diffu1}) and (\ref{eq:diffu2}) this state is obtained when $\dot{u}_j=0$, which leads to the following normalized subradiant state
\begin{equation}
\ket{\psi_-}=r_2\ket{e,g}-r_1\ket{g,e},  \label{eq:subradiantstate}
\end{equation}
which does not decay in time. As the sub-radiant state does not evolve in time, the only its orthogonal state, namely super-radiant state, evolves in time, which reads as
 \begin{equation}
 \ket{\psi_+}=r_1\ket{e,g}+r_2\ket{g,e}.  \label{eq:superradiantstate}
 \end{equation}
Its survival amplitude $\varepsilon(t)$ can be obtained after some manipulations as
 \begin{equation}
\varepsilon(t)\equiv \braket{\psi_+}{\psi_+(t)}=e^{-\kappa t}\left( \cosh{\left(\Omega t/ 2 \right)} +\frac{\kappa}{\Omega}\sinh{\left( \Omega t/ 2\right) }\right),  \label{eq:survivalamplitude}
 \end{equation}
where $\Omega=\sqrt{\kappa^2-4g_{_{T}}^2}$. By introducing $\beta_{\pm}=\braket{\psi_{\pm}}{\psi_0}$, one can obtain  $u_j(t)$ as follow \cite{Maniscalco2008}
\numparts
\begin{eqnarray}
u_{1}(t)&=r_2\beta_-+r_1\varepsilon(t)\beta_+,  \label{eq:u1} \\
u_{2}(t)&=-r_1\beta_-+r_2\varepsilon(t)\beta_+ . \label{eq:u2}
\end{eqnarray}
\endnumparts
As it is seen, the obtained solution is quite exact and we have not used any approximation. We are now in a position to investigate the dynamics of entanglement as measured by concurrence. To begin with, as it is mentioned that there is a decoherence-free state due to interaction of atoms with common environment (see Eq. (\ref{eq:subradiantstate})), there must exist a non-zero stationary value of $C$. We note that, if $t\rightarrow \infty$, then $\varepsilon(t)\rightarrow 0$. So, in the stationary state, $u_1=r_2\beta_-$ and $u_2=-r_1\beta_-$, which leads to a non-zero value of concurrence as
\begin{equation}
C_s=2\left| r_1r_2\right| \left| \beta_-\right| ^2. \label{eq:stationaryconcurrence}
\end{equation}
In the following, we assume that the initial state of system to be such that
\numparts
\begin{eqnarray}
c_{01}&=\sqrt{\frac{1-s}{2}},  \label{eq:c01} \\
c_{02}&=\sqrt{\frac{1+s}{2}}e^{i\varphi},  \label{eq:c01}
\end{eqnarray}
\endnumparts
in which $s$ is the separability parameter with $0\leq s\leq 1$ and $s=\pm 1$ ($s=0$) corresponds to a  separable (maximum entangled) initial state.
In Fig. \ref{fig:stationaryconcurrence} we have plotted the stationary concurrence as a function of the relative coupling constant $r_1$ and the initial separability parameter $s$ for two values of $\varphi$, i.e., $\varphi=0$ and $\pi$. It can clearly be seen that, the separable initial states ($s=\pm 1$) become entangled due to the interaction with the cavity field. In both cases $\varphi=0$ and $\pi$, for $r_1=0$ and $1$ (only the first and second atom interact with cavity field, respectively) the entanglement vanishes as may be expected. This is due to the fact that in this situation, only one atom interacts with the cavity field and there is no correlation between atoms via cavity field. In the case  $\varphi=0$, the maximum stationary entanglement $C_s^{\mathrm{max}}\simeq 0.65$ is achievable for  factorized initial states, i.e., this value is obtained at $r_1=0.5$ for $s=-1$ and at $r_1=0.87$ for $s=+1$. In the case $\varphi=\pi$, the maximum value of  the stationary entanglement $C_s^{\mathrm{max}}=1$ is obtained at $r_1\simeq 0.7$ for the maximum entanglement initial state ($s=0$), which according to Eq. (\ref{eq:subradiantstate}), this maximum is achieved for $\ket{\psi_0}=\ket{\psi_-}$. We point out that the results are independent of $\kappa$. \\
In the following, we intend to investigate the entanglement dynamics of the mentioned system. By introducing two dimensionless parameters $\tau=\kappa t$ and $R=\frac{g_{_{T}}}{\kappa}$, we are able to discuss our results in two regimes, good and bad cavity, i.e. $R\gg 1$ and $R\ll 1$, respectively. Figs. \ref{fig:conbadcavity} and \ref{fig:congoodcavity} show the dynamics of entanglement for bad ($R=0.1$) and good ($R=10$) cavity limits, respectively. We investigate DEM for an initially factorized state ($s=1$) and an initially entangled state ($s=0$) for four values of the coupling parameter $r_1$, namely, $r_1=0.87$, $1/\sqrt 2$, $1$ and $0$. The plots for $r_1=1$ and $r_1=0$ overlap, because they both describe the case in which one atom effectively coupled to the cavity.  We also investigate the influence of phase $\varphi$ on the behaviour of entanglement.

In the case of weak coupling or bad cavity ($R=0.1$) for an initially factorized state ($s=1$), for $r_1=0.87$, $1/\sqrt 2$, the concurrence monotonically increases and reaches its stationary value, $C_s$, whereas, for the case in which only one atom coupled to the cavity, concurrence remains zero as $\tau$ goes on. This is valid, since there is no correlation between atoms in the beginning of interaction and any time. Note that, in this case ($s=1$) these results are independent of phase $\varphi$ (see Figs. \ref{fig:bads10} and \ref{fig:bads1p}), because in this case, $\varphi$ is just a global phase factor. The behaviour of entanglement for an initially entangled state (with $s=0$) is different and as one can see from plots \ref{fig:bads00} and \ref{fig:bads0p}, this behaviour depends on phase $\varphi$. For $\varphi=0$ and $r_1=0.87$, the concurrence first goes to zero before increasing towards $C_s$ (\ref{fig:bads00}), whereas, for $\varphi=\phi$, concurrence decreases monotonically down to its stationary value (\ref{fig:bads0p}). In the case of symmetrical coupling ($r_1=1/\sqrt 2$) with $\varphi=0$ concurrence decays as $\tau$ proceeds and  concurrence vanishes for enough large values of $\tau$. Whereas, for $\varphi=\pi$ the concurrence attains its maximum value (i.e., 1) as $\tau$ goes on. This behaviour is expected, because in this case $\ket{\psi_0}=\ket{\psi_-}$ (see Eq. (\ref{eq:subradiantstate})). For $r_1=0$ and $1$, the DEM is independent of $\varphi$ and as one can see from plots \ref{fig:bads00} and \ref{fig:bads0p}, in this case, concurrence decreases monotonically down to zero. \\
Also, from Fig. \ref{fig:congoodcavity}, an oscillatory behaviour of entanglement is seen nearly for all initial states in the strong coupling. In this case ($R=10$) and for an initially separable state ($s=1$), the results are independent of phase $\varphi$ similar to the cases of weak coupling (Figs. \ref{fig:goods10} and \ref{fig:goods1p}). In this case ($R=10$ and $s=1$), the concurrence increases from zero and oscillates until it reaches its stationary value for  $r_1=0.87$, $1/\sqrt 2$. For  $r_1=0$ and $1$, the concurrence remains zero as $\tau$ goes on. For an initially entangled state ($s=0$) with $\varphi=\pi$, the DEM has an interesting behaviour, i.e., for all values of $r_1$ concurrence falls down from its maximum value and oscillates until it vanishes (see plot \ref{fig:goods00}). \\
The entanglement sudden death is also clearly maybe seen. For $\varphi=\pi$, with $r_1=0$ and $1$, the concurrence has the same behaviour with that of $\varphi=0$ (see Fig. \ref{fig:goods0p}). For $r_1=0.87$, the concurrence falls down from its maximum value and oscillates to its stationary value. In the case of $R=10$ and $s=1$, the concurrence remains at its maximum value (i.e., $1$), since, this situation represents a sub-radiant initial state.

\section{Protecting of Entanglement}
\begin{figure}[htb!]
\centering
\subfigure[\label{fig:good} \ Strong coupling, $R=10$.]{\includegraphics[width=0.4\textwidth]{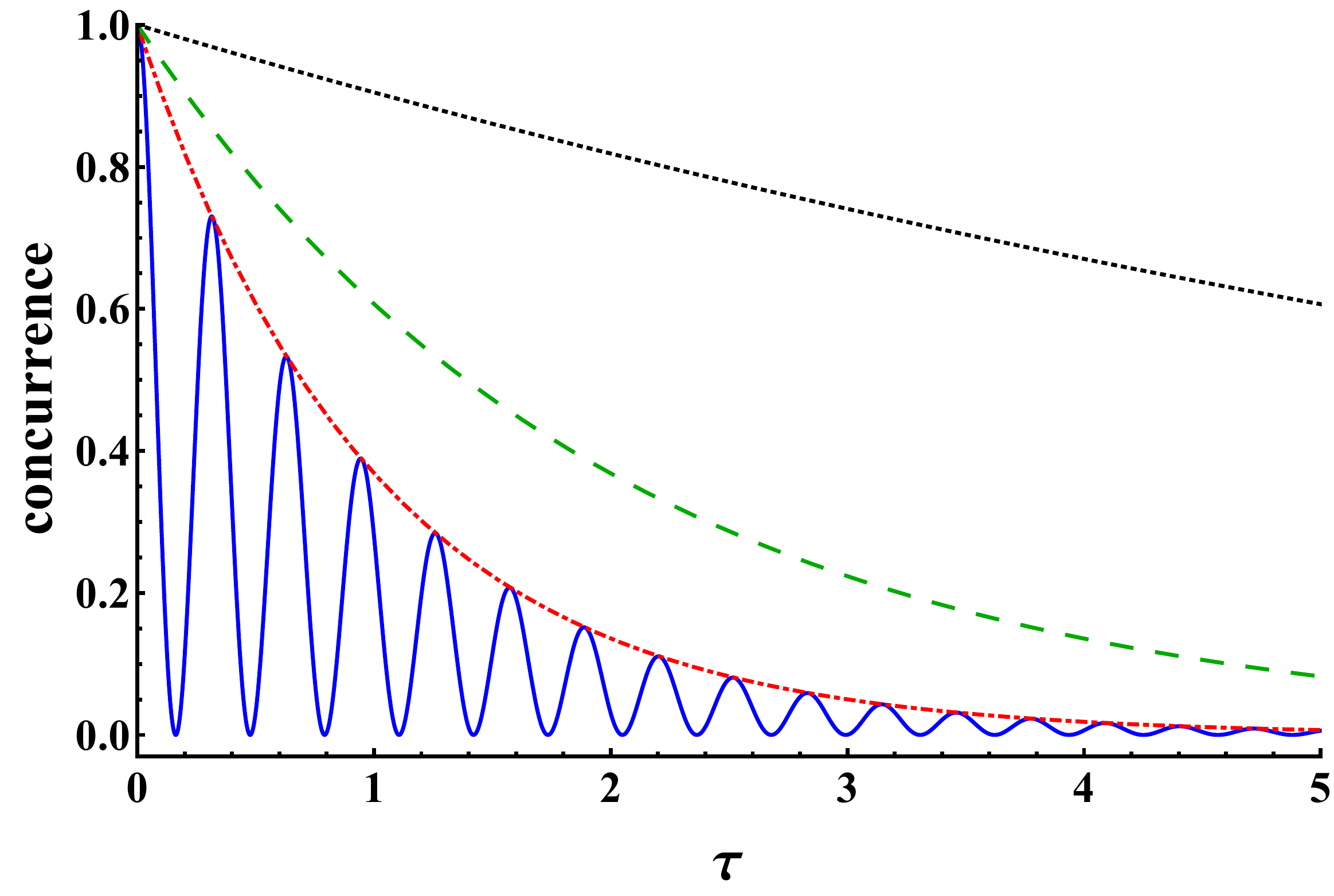}}
  \hspace{0.05\textwidth}
\subfigure[ \label{fig:bad} \ Weak coupling, $R=0.1$.]{\includegraphics[width=0.4\textwidth]{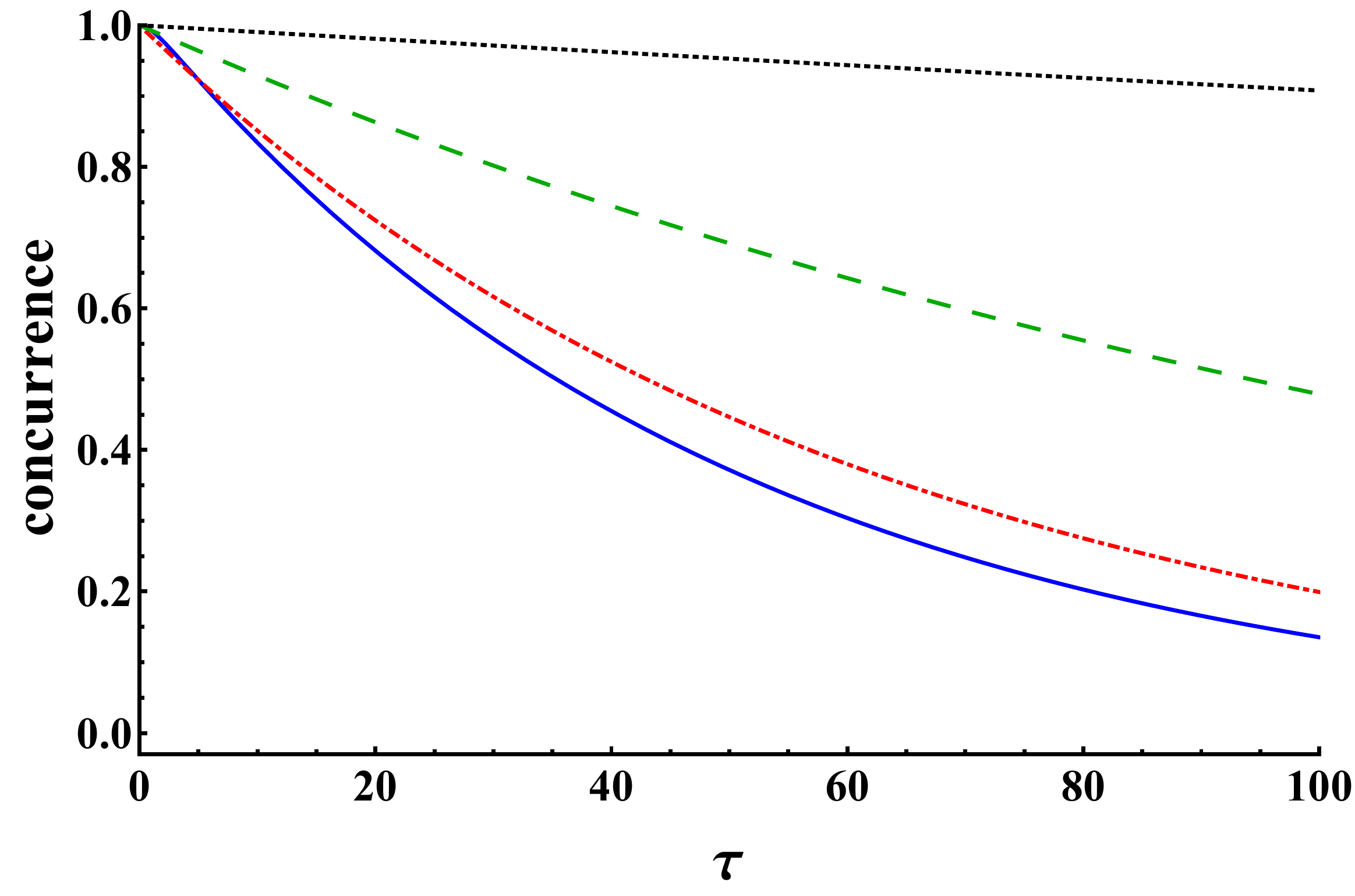}}

\caption{Time evolution of the concurrence for maximally entangled initial state ($s=0$) and symmetric coupling ($r_1=1/\sqrt 2$), in the absence of measurements (solid line) and in the presence of measurements for (a) strong coupling ($R=10$) with intervals $\kappa T=0.01$ (dotted-dashed), $0.005$ (dashed) and $0.001$ (dotted) and (b) weak coupling ($R=0.1$) with intervals $\kappa T=5$ (dotted-dashed), $1$ (dashed) and $0.1$ (dotted).} \label{fig:nconcurrence}
\end{figure}

Consider a special system characterized by Hamiltonian $H$ which its initial state is $\ket{\psi_0}$. The probability of the regarded system being in its initial state is
\begin{equation}
P(t)=\left| \bra{\psi_0}e^{-iHt}\ket{\psi_0}\right|^2. \label{eq:probabilitygeneral}
\end{equation}
The short-time expansion of (\ref{eq:probabilitygeneral}) is then given by $P(t)\simeq 1-\frac{t^2}{\tau_z^2}$, where $\tau_z^{-2}=\bra{\psi_0}H^2\ket{\psi_0}-\bra{\psi_0}H\ket{\psi_0}^2$ is known as Zeno time \cite{Facchi1998}. Let us perform N measurements at time intervals $T=t/N$ in order to check whether the system is still in its initial state. After every measurement, the system is projected back to its initial state and then the temporal evolution starts anew. The survival probability $P^{(N)}(t)$ at the final time $t=NT$ reads
\begin{equation}
P^{(N)}(t)=P(T)^N\simeq\left( 1-\frac{1}{\tau_z^2}\left( \frac{T}{N}\right) ^2\right) ^N\sim 1-\frac{1}{N}\frac{T^2}{\tau_z^2}. \label{eq:zenoeffect}
\end{equation}
It is clear that as $N\rightarrow \infty$, $P^{(N)}(t)\rightarrow 1$. This is the well-known quantum Zeno effect. \\
In our system, we can express the initial state of the system as $\ket{\psi_0}=\beta_+\ket{\psi_+}+\beta_-\ket{\psi_-}$, in which, one can clearly see that the decay of the initial state is directly relevant to the decay of the super-radiant initial state, since the sub-radiant state does not evolve in time. So, it seems logical to find an expression for the survival probability of super-radiant state $\ket{\psi_+}$. To achieve this purpose, we consider the action of a series of nonselective measurements, each performed at time intervals $T$. After measurement, the state $\ket{\psi_-}$ is unaffected but the state $\ket{\psi_+}$ is affected due to the correlation with the surrounding environment. After the first measurement, the surviving probability of $\ket{\psi_+}$ is $\bra{\psi_+}\rho(T)\ket{\psi_+}=\left|\beta_+ \right|^2\varepsilon(T)^2 $. But, for the second measurement, this probability would be $\varepsilon(T)^2$, because after the first measurement, the initial state for the second measurement is $\ket{\psi_+}$, i.e. $\left|\beta_+ \right|^2=1$ and $\left|\beta_- \right|^2=0$. So, after two successive measurements, the probability of the system being in state  $\ket{\psi_+}$ is $P_+^{(2)}(t=2T)=\left|\beta_+ \right|^2\varepsilon(T)^4$. Similarly, after $N$ measurements, the surviving probability is $P_+^{(N)}(t=NT)=\left|\beta_+ \right|^2\varepsilon(T)^{2N}$. After some straightforward  manipulations, this probability can be rewritten as
\begin{equation}
P_+^{(N)}(t)=\left|\beta_+ \right|^2\exp{\left[  -\lambda_z(T)t\right]  },  \label{eq:probabilitynsuperradiant}
\end{equation}
in which $\lambda_z(T)=-\frac{\log{\left[ \varepsilon(T)^2\right] }}{T}$. This probability leads to the surviving amplitude
\begin{equation}
 \varepsilon^{(N)}(t=NT)=\exp{\left[  -\lambda_z(T)t/2\right]}. \label{eq:survivingamplitudenmeasurements}
\end{equation}
It is clear that, in the limit $T\rightarrow 0$ and $N\rightarrow \infty$ with a finite $t=NT$, $\lambda_z(T)\rightarrow 0$ and the decay is completely suppressed. The repeated measurements also affect the temporal evolution of the entanglement. In fact, according to Eqs. (\ref{eq:survivingamplitudenmeasurements}), (\ref{eq:survivalamplitude}), (\ref{eq:u1}) and (\ref{eq:u2}), the modified concurrence at time $t=NT$, after performing $N$ measurements, is given explicitly by
\begin{equation}
C^{(N)}(t)=2\left| \left( \beta_+r_1e^{-\lambda_zt/2}+\beta_-r_2\right)\left( \beta_+r_2e^{-\lambda_zt/2}-\beta_-r_1\right)\right| ,
\label{eq:nconcuurence}
\end{equation}
whose effective dynamics now depends on $T$. \\
 Fig. \ref{fig:nconcurrence} illustrates the time evolution of the concurrence in the absence (solid lines) and in the presence of the projective measurements performed at various intervals $T$ for an initially maximal entangled state ($s=0$) and symmetric coupling ($r_1=1/\sqrt 2$) in the strong and weak coupling regimes (left and right plots, respectively) with $\varphi=0$. It should be noticed that we do not consider the case $\varphi=\pi$ (with $s=0$ and $r_1=1/\sqrt 2$), because this case represents the sub-radiant state which does not evolve in time. In both coupling regimes, the presence of measurements, suppresses the decay of the concurrence. By comparing Figs. \ref{fig:goods00} and \ref{fig:good}, one can clearly observe that, the entanglement sudden death is completely disappeared due to the repeated measurements. On the other hand, by decreasing the time interval between  the measurements, the concurrence remains closer and closer to its initial value. Accordingly, we could be able to protect the entanglement stored in the mentioned system from the dissipation by quantum Zeno effect. The amount of protecting is directly depended on the time interval between successive measurements and consequently the number of measurements, $N$, with a finite time $t=NT$.

\section{Conclusion}
To summarize, we have introduced a system containing two two-level atoms interacting with a common dissipative cavity. We used the Gardiner-Collett Hamiltonian to describe the dissipation of the cavity. With the help of Fano's method, we showed that, this system can be reduced to a system in which the two atoms dissipate in a common heat bath. Then, by considering the general form of the wave function of the entire system, we solved the time dependent Schr\"{o}dinger equation and obtained the exact analytical solution of the state vector of the system.
 
We showed that, there exists a stationary state which does not evolve in time. The surprising aspect is that an initially separable state becomes entangled due to the interaction with environment and this entanglement remains as time goes on. Remarkably, this phenomenon happens even without any interaction between subsystems (see Eq. (\ref{eq:finalhamiltonian})). Actually, it can be said that the environment can provide an indirect interaction between otherwise decoupled qubits and therefore establishes entanglement between them.

We also investigated the dynamics of entanglement in both weak and strong coupling regimes in details and showed that the behaviour of entanglement depends on the separability parameter, the relative coupling constant between atoms and cavity field and the phase $\varphi$. For a strong coupling, the concurrence has an oscillatory behaviour. The quantum sudden death is seen for the case in which the atoms are initially in  maximum entangled state in the strong coupling regime (see Fig. \ref{fig:goods00}). We also introduced the quantum Zeno effect as a method to preserve entanglement. The results showed that, for both good and bad coupling regimes, the presence of repeated measurements quenches the decay of entanglement and the entanglement sudden death in not seen no longer. As the time intervals between measurement decrease, the concurrence remains closer to its initial value.

 \renewcommand{\theequation}{A.\arabic{equation}}
 \setcounter{equation}{0}  
  \section*{Appendix. Diagonilzation of the $H_{\mathrm{FS}}$.}
 In this appendix, we intend to diagonalize the Hamiltonian (\ref{eq:HFE}). To begin with, let us define the dressed operator:
 \begin{equation}
 \hat{A}(\omega)=\alpha(\omega)\hat{a}+\int_0^{\infty}\! \beta(\omega,\eta)\hat{B}(\eta) \, \mathrm{d}\eta,
 \label{eq:dressedoperatorA}
 \end{equation}
 in terms of which the Hamiltonian (\ref{eq:HFE}) is diagonal

\begin{eqnarray}
\left[ \hat{A}(\omega),\hat{H}_{\mathrm{FE}}\right] =\omega\hat{A}(\omega) \label{eq:digonalizationconditionA1} \\
\hat{H}_{\mathrm{FE}} =\int_0^{\infty}\! \omega \hat{A}^{\dagger}(\omega)\hat{A}(\omega) \, \mathrm{d}\omega, \label{eq:digonalizationconditionA2}
\end{eqnarray}

 and such that $\hat{A}(\omega)$ be an annihilation operator
 \begin{equation}
 \left[ \hat{A}(\omega),\hat{A}^{\dagger}(\omega^{'})\right]=\delta(\omega-\omega^{'}).
 \label{eq:commutationrelationA}
 \end{equation}
 There is no need to check the other commutation relation, i.e. $\left[ \hat{A}(\omega),\hat{A}(\omega^{'})\right]=0$, as it is trivially satisfied by definition of $\hat{A}(\omega)$. So (\ref{eq:dressedoperatorA}), (\ref{eq:digonalizationconditionA2}) and (\ref{eq:commutationrelationA}) are enough to define $\hat{A}(\omega)$ uniquely except for a global phase factor. From (\ref{eq:dressedoperatorA}) and (\ref{eq:digonalizationconditionA2}), we obtain the following system of coupled equations

 \begin{eqnarray}
\omega_c\alpha(\omega)+G\int_0^{\infty}\! \beta(\omega,\eta) \, \mathrm{d}\eta=\omega\alpha(\omega), \label{eq:firstA} \\
\eta\beta(\omega,\eta)+G\alpha(\omega)= \omega\beta(\omega,\eta). \label{eq:secondA}
 \end{eqnarray}
 Solving (\ref{eq:secondA}) for $\beta(\omega,\eta)$, we find that
 \begin{equation}
 \beta(\omega,\eta)=G\alpha(\omega)P\frac{1}{\omega-\eta}+z(\omega)\alpha(\omega)\delta(\omega-\eta). \label{eq:beta1A}
 \end{equation}
 To determine the function $z(\omega)$, we substitute ($\ref{eq:beta1A}$) into (\ref{eq:firstA}), by which we then arrive at
 \begin{equation}
 z(\omega)=\frac{\omega-\omega_c}{G}. \label{eq:zA}
 \end{equation}
 Thus,
 \begin{equation}
 \beta(\omega,\eta)=G\alpha(\omega)P\frac{1}{\omega-\eta}+\frac{\omega-\omega_c}{G}\alpha(\omega)\delta(\omega-\eta). \label{eq:betaA}
 \end{equation}
 In order to determine $\alpha(\omega)$, we substitute (\ref{eq:betaA}) into (\ref{eq:commutationrelationA}) which arrives us at

 \begin{eqnarray}
   &&\hspace{-25mm}\alpha^{*}(\omega^{'})\alpha(\omega) + G^2\alpha^{*}(\omega^{'})\alpha(\omega)\int_{-\infty}^{\infty}\! P\frac{1}{\omega^{'}-\eta}P\frac{1}{\omega-\eta} \, \mathrm{d}\eta + \alpha^{*}(\omega^{'})\alpha(\omega)\frac{\omega-\omega_c}{\omega^{'}-\omega} \nonumber \\
    &&\hspace{+2mm}+\alpha^{*}(\omega^{'})\alpha(\omega)\frac{\omega^{'}-\omega_c}{\omega-\omega^{'}}+\left( \frac{\omega-\omega_c}{G}\right)^2\alpha^{*}(\omega^{'})\alpha(\omega)\delta(\omega-\omega^{'}) \nonumber \\
    &&\hspace{+2mm}=\delta(\omega-\omega^{'}).
   \label{eq:alphaeqA}
  \end{eqnarray}
Using the identity \cite{Dutra2005}
\begin{equation}
\int_{-\infty}^{\infty}\! P\frac{1}{\omega^{'}-\eta}P\frac{1}{\omega-\eta} \, \mathrm{d}\eta=\pi^2\delta(\omega-\omega^{'})
\label{eq:principlevalue}
\end{equation}
Eq. (\ref{eq:alphaeqA}) becomes
\begin{equation}
\left[ \vphantom{\left( \frac{\omega-\omega_c}{G}\right)^2} \pi^2G^2\alpha^{*}(\omega^{'})\alpha(\omega) \right.+ \left.  \left( \frac{\omega-\omega_c}{G}\right)^2\alpha^{*}(\omega^{'})\alpha(\omega)\right] \delta(\omega-\omega^{'}) =\delta(\omega-\omega^{'}).
\end{equation}
From this relation we find
\begin{equation}
\left|\alpha(\omega)\right|^2=\frac{G^2}{\left(\omega-\omega_c\right)^2 +\pi^2G^4 }.
\label{eq:normalphaA}
\end{equation}
Choosing the arbitrary phase, we are lead to $(G=\sqrt{\frac{\kappa}{\pi}})$
\begin{equation}
\alpha(\omega)=\frac{\sqrt{\frac{\kappa}{\pi}}}{\omega-\omega_c +i\kappa }.
\label{eq:mailalphaA}
\end{equation}
We can also express the bosonic operators $\hat{a}$ in terms of the introduced dressed operator. We assume that, $\hat{b}$ can be expressed as a linear combination of the $\hat{A}(\omega)$,
\begin{equation}
\hat{a}=\int_0^{\infty}\! \zeta(\omega)\hat{A}(\omega) \, \mathrm{d}\omega.
\label{eq:linearcombinationA}
\end{equation}
Now, we calculate the commutator $\left[ \hat{a},\hat{A}^{\dagger}(\omega)\right]$ in two ways. First, we substitute $\hat{a}$ as linear combination of  $\hat{A}(\omega)$ in (\ref{eq:linearcombinationA}). Second, we substitute the Hermition conjugate of (\ref{eq:dressedoperatorA}) for $\hat{A}^{\dagger}(\omega)$. By comparing these two results, we find that $\zeta(\omega)=\alpha^{*}(\omega)$.

\end{document}